\documentclass[sigconf,nonacm]{acmart} 

\AtBeginDocument{%
  }

\usepackage{tabularx}
\usepackage{multirow}
\usepackage{enumitem}
\usepackage{tikz}
\usepackage{pifont}
\usepackage{subfig}
\usepackage{siunitx}  
\usepackage{makecell} 
\usepackage{algorithm}
\usepackage{algpseudocode}

\setlist[itemize]{leftmargin=1.25em,itemsep=0pt,parsep=1.5pt,topsep=0pt,partopsep=0pt}
\setlist[enumerate]{leftmargin=1.25em,itemsep=0pt,parsep=1.5pt,topsep=0pt,partopsep=0pt}

\definecolor{posblue}{RGB}{0, 176, 240} 
\definecolor{negred}{RGB}{255, 0, 0}    
\definecolor{zebra}{gray}{0.95}         

\setcopyright{none}
\renewcommand\footnotetextcopyrightpermission[1]{}
\acmConference[ICMR '26]{International Conference on Multimedia Retrieval}{June 16--19, 2026}{Amsterdam, Netherlands}
\acmBooktitle{International Conference on Multimedia Retrieval (ICMR '26), June 16--19, 2026, Amsterdam, Netherlands}

\begin{document}

\title{From Physics to Representation: Audio Learning with Synthetic Pre-training via Procedural Generation}


\author{Fengrui Liu}
\orcid{0009-0006-5646-990X}
\affiliation{%
  \institution{East China Normal University}
  \city{Shanghai}
  \country{China}
}
\email{FengruiLiu_0419@stu.ecnu.edu.cn}

\author{Ruiyang Huang}
\orcid{0009-0007-0178-8572}
\affiliation{%
  \institution{Southeast University}
  \city{Nanjing}
  \state{Jiangsu}
  \country{China}
}
\email{ryhuang_572@seu.edu.cn}

\author{Qijian Zheng}
\affiliation{%
  \institution{Fudan University}
  \city{Shanghai}
  \country{China}
}
\email{qjzheng25@m.fudan.edu.cn}

\author{Yuanfang Wang}
\affiliation{%
  \institution{Shanghai Jiao Tong University}
  \city{Shanghai}
  \country{China}
}
\email{zhaomc911@sjtu.edu.cn}

\author{Feng Liu}
\authornote{Corresponding author.}
\orcid{0000-0002-5289-5761}
\affiliation{%
  \institution{Shanghai Jiao Tong University}
  \city{Shanghai}
  \country{China}
}
\email{liu.feng@sjtu.edu.cn}


\renewcommand{\shortauthors}{Liu et al.}

\begin{abstract}
Self-supervised learning advances audio representation for multimedia analysis. However, prevailing data-centric approaches rely on massive real-world corpora, increasing training costs, curation burdens, and privacy barriers. To address this, we present AudioPG, a procedural synthesis framework eliminating real audio recordings during pre-training. AudioPG trains a Transformer-based masked autoencoder on waveforms generated on-the-fly from basic acoustic primitives and composition rules. The encoder transfers effectively to real audio benchmarks, achieving 90.60\% accuracy on ESC-50, 0.546 mAP on FSD50K, 88.17\% on UrbanSound8K, and 97.03\% on Speech Commands V2. Notably, pre-training completes in under 20 minutes on a single GPU. Latent space analysis reveals physical factors, including fundamental frequency and relative intensity, emerge in orthogonal subspaces, making representations linearly decodable. These results establish procedural synthesis as an efficient, interpretable pre-training signal when large-scale corpora are unavailable. Our code is available at: https://github.com/Freyliu0516/audioPG.
\end{abstract}



\maketitle

\begin{figure}[t]
  \captionsetup{skip=0.5em}
  \centering
  \includegraphics[width=\columnwidth]{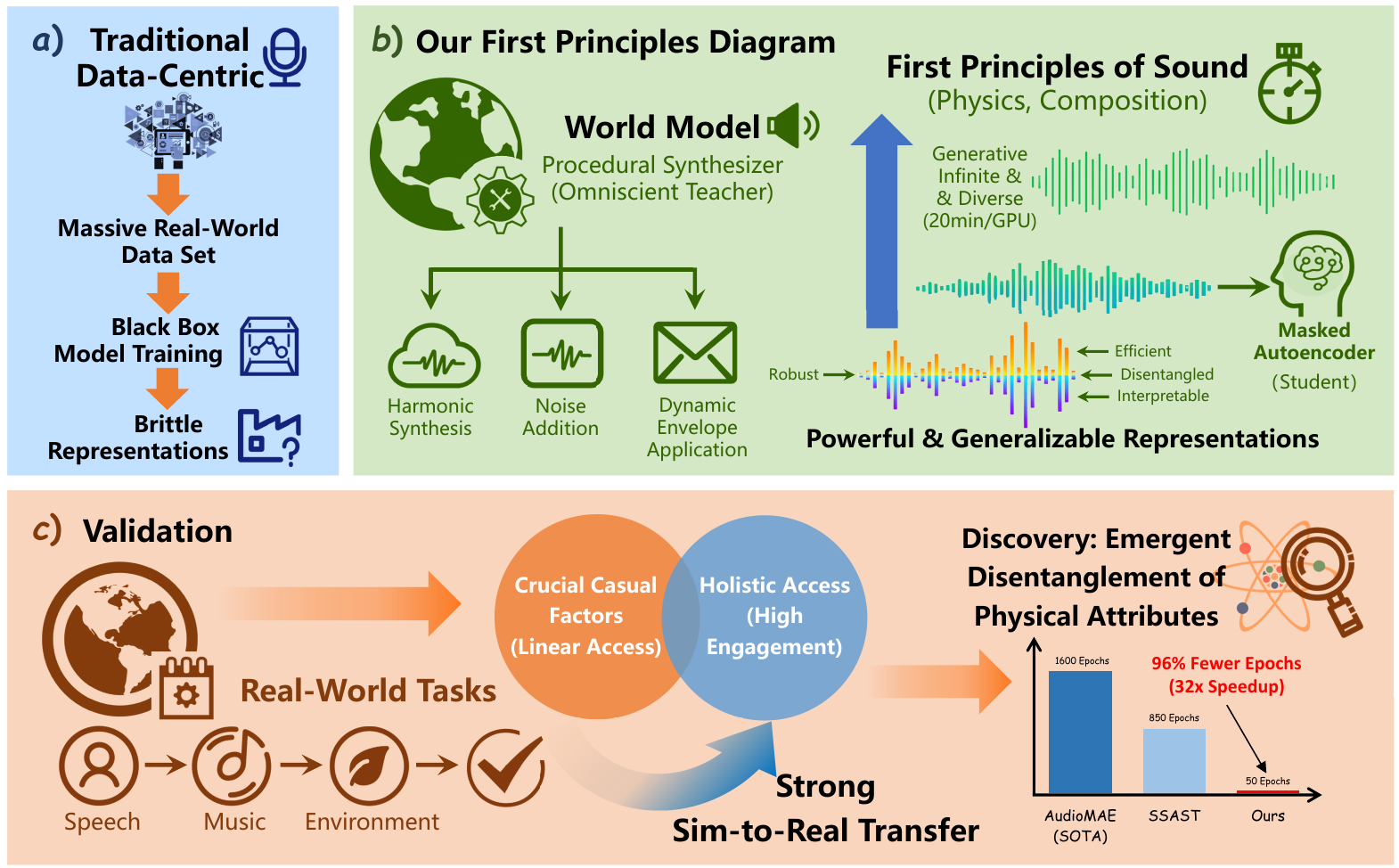}
  \caption{Overview of the proposed AudioPG framework.}
  \label{fig:fp_paradigm}
\end{figure}
\section{Introduction}
\label{sec:intro}

Self-supervised learning (SSL) has become a common approach for learning audio representations from unlabeled recordings~\cite{baevski2020wav2vec,ssast,niizumi2021byol}. 
In parallel, supervised pre-training on large audio corpora can also yield transferable embeddings including PANNs~\cite{kong2020panns}, and speech-oriented SSL models including HuBERT~\cite{hsu2021hubert} have shown strong downstream performance. 
Despite this progress, most current pipelines still rely on large collections of real recordings including AudioSet~\cite{gemmeke2017audioset} or LibriSpeech~\cite{panayotov2015librispeech}. 
Building and using such corpora is costly, which involves significant computational and data curation burdens, and is often difficult in settings limited by privacy, licensing, or the availability of specific sound events. 
A second issue is that objectives learned from natural recordings are largely determined by the correlations present in the training distribution. 
In practice, representations may capture dataset or recording condition cues that are useful for optimization but provide limited control over, or insight into, the underlying generative factors of sound~\cite{bengio2013representation,geirhos2020shortcut}. 
These observations motivate a different kind of pre-training signal, which can be scaled easily, varied systematically, and tied to an explicit physical construction of audio. 
In this work, we investigate whether general-purpose audio representations can be learned without any real recordings by exploiting physical principles of sound synthesis. 
We introduce AudioPG, which pre-trains a masked autoencoder on waveforms generated on-the-fly by a lightweight procedural synthesizer. 
The synthesizer is specified by a small set of acoustic primitives together with composition rules~\cite{farnell2010designing}, allowing us to vary timbre, temporal dynamics, and spectral shaping through explicit parameters.

As illustrated in Fig.~\ref{fig:fp_paradigm}, our setting departs from the standard data-centric pipeline by replacing large real-audio corpora with procedurally generated signals whose generative factors are directly parameterized.
This offers a controllable training curriculum and supports the analysis of how physically meaningful attributes are reflected in the learned representation.
The synthesizer produces waveforms by superposing basic building blocks, including harmonic additive synthesis, frequency modulation, broadband pulse trains, transient bursts, multi-event ADSR envelopes, spectral damping via low-pass filtering, and background noise.
Peak normalization is applied after synthesis to remove absolute gain; relative intensity is varied via signal-to-noise ratio settings.
Diversity is obtained by sampling parameters from designed distributions that span a wide range of timbres, temporal patterns, and spectral shapes.
On top of this curriculum, we train a Transformer-based masked autoencoder to reconstruct log-Mel spectrograms under heavy masking.
With a masking ratio of 75\%, the model must infer the missing time and frequency structure from sparse visible patches, encouraging it to leverage the compositional regularities induced by the synthesis process.
The remainder of this paper is organized as follows. Section~\ref{sec:related} reviews related work. Section~\ref{sec:methodology} details our proposed AudioPG framework.
Section~\ref{sec:experiments} presents the experimental evaluation, and Section~\ref{sec:discussion} discusses our empirical results. Finally, we conclude our work in Section~\ref{sec:conclusion}.
The key contributions of this work are listed as follows:
\begin{itemize}
    \item We successfully verify the feasibility of audio pre-training completely detached from real data.
    \item The resulting encoder demonstrates strong cross-domain transfer capabilities across multiple real-world benchmarks including ESC-50 \cite{piczak2015esc}, UrbanSound8K \cite{salamon2014dataset}, FSD50K \cite{fonseca2022fsd50k}, and Speech Commands V2 \cite{warden2018speech}.
    \item An in-depth analysis of the model latent space proves that this physics-based reconstruction task can prompt the feature space to spontaneously decouple physical attributes including frequency and relative intensity, providing a new path for audio representation learning that balances efficiency and interpretability \cite{bengio2013representation}.
\end{itemize}
\section{Related Works}
\label{sec:related}

\noindent\textbf{Data-Driven Audio Self-Supervision.}
Self-supervised learning (SSL) targets transferable audio representations from unlabeled recordings. Early efforts relied on large-scale supervised pre-training \cite{kong2020panns, gemmeke2017audio, li2026sepprune} before transitioning to instance-level discrimination with contrastive objectives \cite{oord2018cpc, schneider2019wav2vec, baevski2020wav2vec}. Subsequent developments incorporated future observation prediction \cite{lian19b_interspeech}, domain-tailored augmentations \cite{saeed2021cola}, mixing-based regularization including mixup \cite{zhang2018mixup}, and self-distillation \cite{niizumi2021byol, grill2020byol, li2024atst}. Inspired by natural language processing \cite{devlin2019bert}, masked prediction objectives became prominent. This line of research encompasses methods relying on discrete units \cite{baevski2020vqwav2vec}, iterative clustering \cite{hsu2021hubert}, continuous reconstruction \cite{liu2021tera}, denoising \cite{chen2021wavlm}, teacher-guided latent targets \cite{baevski2022data2vec}, unpaired textual data alignment \cite{zhang2024speechlm}, and cross-utterance context modeling \cite{cui2025context}. Transformer architectures treating time and frequency patches as tokens currently dominate these masked autoencoding frameworks \cite{gong2021ast, ssast, he2022masked, mae_ast, huang2022listen, msm_mae, maskspec, gong2023contrastive, 11094565}, with performance depending on normalization and attention design \cite{xiong2020layer, vaswani2017attention}. While effective, these data-centric approaches require massive real-audio corpora and often encode dataset or recording condition biases \cite{whetten2026study}, motivating the exploration of controllable alternative pre-training signals.

\noindent\textbf{Procedural Synthesis and Representation Quality}
Procedural audio synthesis generates sound using parameterized primitives reflecting physical sound production \cite{farnell2010designing}. In machine learning pipelines, synthetic audio traditionally serves as augmentation \cite{salamon2017scaper} or domain randomization \cite{tobin2017domain}. Other directions incorporate signal-processing structures via differentiable modules \cite{engel2020ddsp} or utilize diffusion models for enhancement \cite{welker2022sgmse, richter2023diffusion}. Recent studies indicate that pre-training on non-acoustic synthetic patterns including visual fractals can transfer to spectrogram modeling \cite{ishikawa2025pre}. In contrast, our approach utilizes knowledge-driven procedural generation with explicitly parameterized acoustic factors. This relates to the broader goal of capturing task-relevant structures while separating independent generative factors \cite{bengio2013representation}. While explicit regularization \cite{higgins2017beta, kim2018factorvae} or inductive biases \cite{locatello2019challenging} are typically required to identify such factors, procedural generation provides data with known parameters. This structure allows for testing whether standard reconstruction objectives yield latents correlating with physically meaningful attributes without auxiliary losses, thereby connecting downstream transfer performance to structured representation quality.
\begin{figure*}[t]
   \captionsetup{skip=-0.05em}
  \centering
  \includegraphics[width=\textwidth]{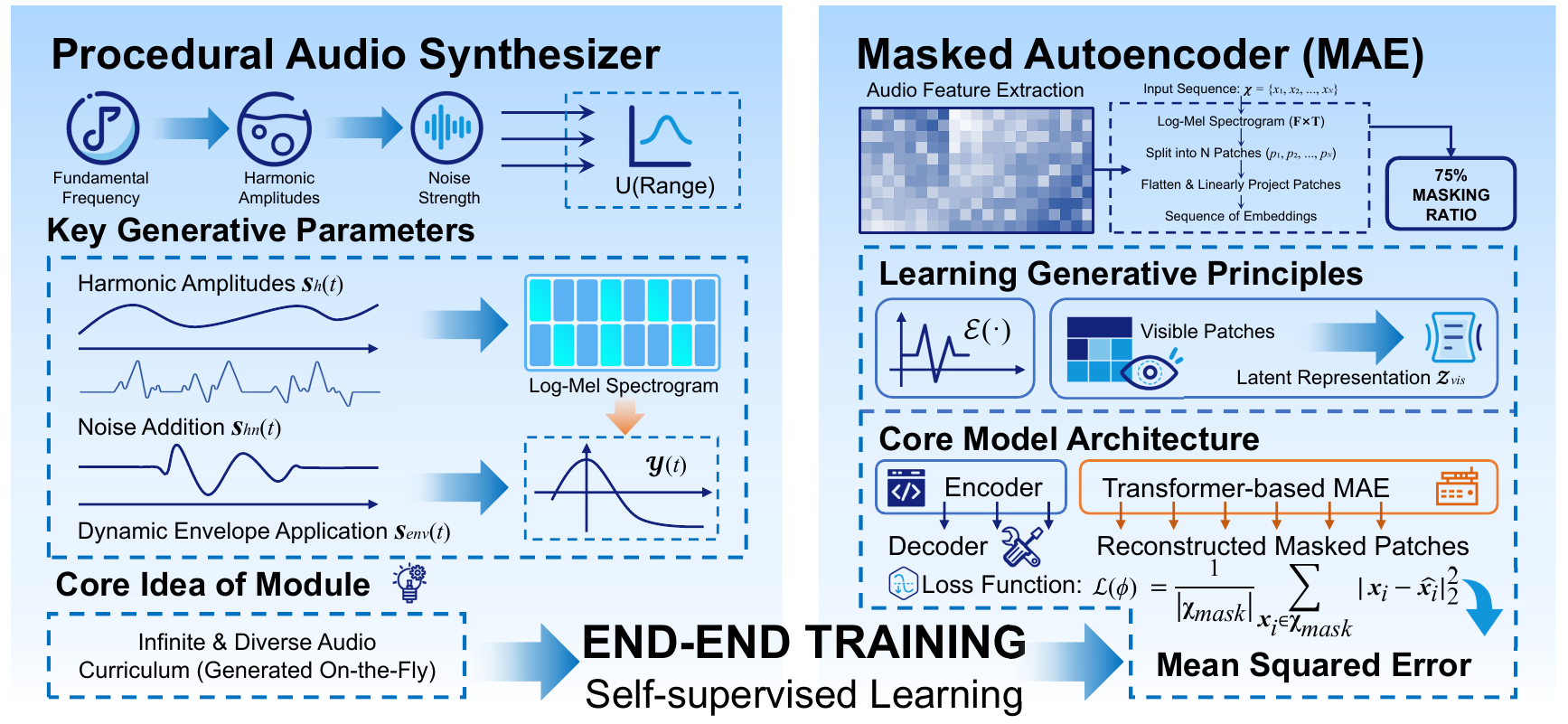}
  \caption{The framework of AudioPG.
(Left) The Procedural Audio Synthesizer generates an unbounded curriculum of acoustic events using parameterized primitives including harmonic oscillation, stochastic modulation, and transient bursts.
(Right) A Transformer-based MAE is trained to reconstruct missing time and frequency patches from masked spectrograms.
Pre-training on procedurally generated signals provides explicit control over generative factors and supports analysis of how compositional structure is encoded, without using real recordings during pre-training.}
  \label{fig:overview}
\end{figure*}

\section{Methodology}
\label{sec:methodology}

\subsection{Procedural Audio Synthesizer}
\label{ssec:synthesizer}

As shown in fig\ref{fig:overview} , we define a parametric generator $\mathcal{G}(\theta)$ mapping a parameter set $\theta$ to a waveform $y(t)$. The generation process follows a source and filter model where an excitation source is formed by composing acoustic primitives with temporal envelopes, followed by a global spectral shaping stage to simulate damping \cite{farnell2010designing}. Given sampled parameters $\theta$, the generator constructs a raw signal $\tilde{y}(t)$ as the sum of an event-based source term and an additive noise floor, followed by peak normalization:
\begin{equation}
    y(t) = \frac{\tilde{y}(t)}{\max_{\tau}|\tilde{y}(\tau)| + \epsilon}
    \quad \text{where } \tilde{y}(t) = s_{\text{src}}(t) + \lambda_n \eta(t)
    \label{eq:normalization} 
\end{equation}
Here, $\eta(t)$ represents background noise and $\lambda_n$ controls its intensity. The source term $s_{\text{src}}(t)$ is a superposition of $N_e$ acoustic events, each shaped by an ADSR envelope and an optional transient component, followed by spectral damping:
\begin{equation}
    s_{\text{src}}(t) = \left[ \sum_{i=1}^{N_e} A_i \cdot \mathcal{E}_i(t) \cdot \left( \Psi_{\text{osc}}^{(i)}(t) + \Psi_{\text{trans}}^{(i)}(t) \right) \right] * h_{\text{damp}}(t)
    \label{eq:source_model}
\end{equation}
where $A_i$ is the event amplitude, $\mathcal{E}_i(t)$ is an ADSR envelope, $\Psi_{\text{osc}}^{(i)}(t)$ is the tonal excitation, $\Psi_{\text{trans}}^{(i)}(t)$ is a transient burst, and $h_{\text{damp}}(t)$ is a low-pass damping filter. Because Equation \ref{eq:normalization} removes absolute gain, variations in $A_i$ alter the effective signal-to-noise ratio rather than the absolute signal level.

To instantiate $\Psi_{\text{osc}}$, we utilize three excitation modes including harmonic additive synthesis, frequency modulation, and broadband pulse synthesis. The harmonic additive synthesis constructs a sum of $K$ harmonics with a fundamental frequency $f_0$ and a power-law roll-off $\gamma$. The frequency modulation mode uses a two-operator form parameterized by a modulation index $I$ and a carrier-modulator ratio $r$. Broadband pulse synthesis employs geometric waveforms including sawtooth and square waves to produce dense harmonic content. To introduce nonstationary structures, we sample $N_e \in [1,5]$ events and assign them random temporal positions. Each event is modulated by an ADSR envelope. Impulsive onsets are simulated by adding short noise bursts with a probability $p_b$. Finally, frequency-dependent attenuation is modeled with a stochastic low-pass biquad filter $h_{\text{damp}}(t)$ applied with probability $p_f$ and a cutoff frequency $f_c$. The full parameterization is detailed in supplementary material. By continuously sampling these variables from predefined distributions, the generator produces an unbounded curriculum encompassing diverse timbres, temporal patterns, and spectral shapes. This systematic variation prevents the network from memorizing specific instances, encouraging it to capture the underlying compositional regularities necessary for robust generalization.

\subsection{Masked Autoencoder Learning Framework}
\label{ssec:mae}

We adopt a masked autoencoding objective \cite{he2022masked, huang2022listen} on log-Mel spectrograms computed from the synthesized waveforms. Given $y(t)$, we compute a log-Mel representation $\mathbf{X} \in \mathbb{R}^{T \times F}$ with $T=1024$ frames and $F=128$ Mel bins, applying global standardization using dataset-level mean and standard deviation statistics. The matrix $\mathbf{X}$ is divided into non-overlapping $16 \times 16$ patches. We apply a random masking strategy with a ratio of $\rho=75\%$, retaining only the visible subset of patches. A Transformer encoder \cite{vaswani2017attention} processes the visible patches, and a decoder reconstructs the full spectrogram by filling in masked locations using mask tokens. The network is trained end-to-end to minimize the mean squared error between the model predictions and the normalized masked patches.
\section{Experiments}
\label{sec:experiments}

\subsection{Experimental Setup}
\label{ssec:setup}

We evaluate transfer performance on four real-audio benchmarks spanning environmental sounds, urban sounds, open-domain audio events, and keyword spotting.
ESC-50\cite{piczak2015esc} contains 2,000 clips from 50 classes, where we follow the standard cross-validation protocol and report mean accuracy across folds.
UrbanSound8K\cite{salamon2014dataset} contains 8,732 clips from 10 classes, where we use the official split and report accuracy.
FSD50K\cite{fonseca2022fsd50k} contains 51,197 open-domain clips with multi-label annotations and variable duration, where we use the standard development and evaluation split and report mAP.
FSD50K is multi-label and often polyphonic, which probes transfer from our predominantly monophonic procedural curriculum to real-world mixtures.
Speech Commands V2\cite{warden2018speech} contains 105,829 one-second utterances over 35 keywords, where we use the standard splits and report accuracy.
\begin{table*}[thbp]
\centering
 \captionsetup{skip=2.5pt}
\caption{Fine-tuning performance comparison on four benchmarks.}
\label{tab:overall_results}
\fontsize{9pt}{10.8pt}\selectfont
\setlength{\aboverulesep}{0pt}
\setlength{\belowrulesep}{0pt}
\renewcommand{\arraystretch}{1.05}
\setlength{\tabcolsep}{7.5pt}
\begin{tabularx}{0.98\textwidth}{Xlcccc}
\toprule
\textbf{Method} & \textbf{Pre-train Data} & {\makecell[c]{\textbf{ESC-50}\\\textbf{Acc.↑ (\%)}}} & {\makecell[c]{\textbf{US8K}\\\textbf{Acc.↑ (\%)}}} & {\makecell[c]{\textbf{SCv2}\\\textbf{Acc.↑
(\%)}}} & {\makecell[c]{\textbf{FSD50K}\\\textbf{mAP↑}}} \\
\midrule
\multicolumn{6}{l}{\quad\textit{Traditional Baselines}} \\
Random Guess (1/K) & -- & 2.00 & 10.00 & 2.86 & {--} \\
MFCC + Random Forest\cite{piczak2015esc} & -- & 44.3 & {--} & {--} & {--} \\
MFCC + SVM\cite{piczak2015esc} & -- & 39.6 & {--} & {--} & {--} \\
SKM\cite{salamon2017dcnn} & -- & {--} & 74.0 & {--} & {--} \\
VGG-like\cite{fonseca2022fsd50k} & -- & {--} & {--} & {--} & 0.434 \\
\midrule
\multicolumn{6}{l}{\quad\textit{Supervised Baseline}} \\
Supervised (Scratch) & -- & 54.00 & 75.34 & 96.30 & 0.398 \\
\midrule
\multicolumn{6}{l}{\quad\textit{ImageNet Initialization}} \\
AST-S\cite{ast} & IN & 88.7 & {--} & 98.11 & {--} \\
ViT-B (ImageNet SL)\cite{ishikawa2025pre} & IN-1k & 
87.0 & 77.6 & {--} & 0.573 \\
\midrule
\multicolumn{6}{l}{\quad\textit{SOTA with Real-Audio Pre-training}} \\
PANNs (CNN14)\cite{kong2020panns} & AudioSet (Sup.) & 94.7 & 87.4 & 96.9 & 0.431 \\
Wav2Vec 2.0 (Base)\cite{baevski2020wav2vec} & LibriSpeech & 75.5 & {--} & 96.2 & 0.320 \\
BYOL-A\cite{niizumi2021byol} & AudioSet & 84.2 & 84.8 & {--} & 0.395 \\
SSAST\cite{ssast} & AS+LS & 88.8 & 86.5 & 98.0 & 0.465 \\
MAE-AST\cite{mae_ast} & AS+LS & 90.0 & 88.4 & 95.8 & 0.482 \\
MaskSpec\cite{maskspec} & AS & 90.7 & 89.6 & 97.7 & 0.503 \\
MSM-MAE\cite{msm_mae} & AS & 94.0 & 88.1 & {--} & 0.490 \\
AST (Single)\cite{ast} & IN+AS & 95.6 & 89.8 
& 98.1 & 0.514 \\
PaSST (IN+AS)\cite{passt} & IN+AS & 96.8 & 90.1 & {--} & \textbf{0.653} \\
BEATs (iter3+)\cite{beats} & AS & \textbf{98.1} & \textbf{91.1} & 98.1 & 0.562 \\
EAT (2024)\cite{chen2024eat} & AS & 95.9 & {--} & \textbf{98.3} & {--} \\
OpenBEATs (2025)\cite{bharadwaj2025openbeats} & AS & 95.8 & {--} & {--} & 0.575 \\
\midrule
\multicolumn{6}{l}{\textit{SOTA with Synthetic Pre-training}} \\
FDSL (exF-21k)\cite{kataoka2022fdsl,ishikawa2025pre} & Syn.
Images & 83.7 & {--} & {--} & {--} \\
FDSL (VA-21k)\cite{takashima2023visualatoms,ishikawa2025pre} & Syn.
Images & 68.1 & {--} & {--} & {--} \\
MAE (VA-1k)\cite{takashima2023visualatoms,ishikawa2025pre} & Syn.
Images & 79.5 & {--} & {--} & {--} \\
Ishikawa \textit{et al.} (FractalDB1k)\cite{ishikawa2025pre} & Syn.
Images & 13.6 & {--} & {--} & {--} \\
Ishikawa \textit{et al.} (Shaders1k)\cite{ishikawa2025pre} & Syn.
Images & 87.3 & 78.3 & 96.8 & 0.563 \\
\textbf{AudioPG (Ours)} & \textbf{Proc.
Audio} & \textbf{90.60(±2.55)} & \textbf{88.17} & \textbf{97.03} & \textbf{0.546} \\
\bottomrule
\end{tabularx}
\begin{minipage}{0.98\textwidth}
\vspace{0.25em}
\footnotesize \textit{Note:} AudioPG uses 0 real audio during pre-training.
Reported results for prior work follow the corresponding papers.
\end{minipage}
\end{table*}

All audio is resampled to 16 kHz and represented as 128-bin log-Mel filterbank features computed with a 25 ms Hamming window and a 10 ms hop.
Inputs are matched to the pre-training length by central cropping or loop-padding to 10.24 s prior to patchification.
Unless stated otherwise, results are reported under full fine-tuning, where the encoder is initialized from the pre-trained weights and updated end-to-end with the task loss.
For analyses that operate on frozen representations, the encoder is fixed and only a lightweight classifier is trained on top of the frozen features.
For linear probing, we train a lightweight classifier on top of frozen encoder features, where feature aggregation follows the downstream fine-tuning implementation.
We compare against three types of references. The first involves internal controls reproduced locally, including random guess, a raw log-Mel baseline trained without a pre-trained encoder, and a ViT-Base model trained from scratch on the downstream labels.
The second involves compute-matched real-data references, where the same AudioMAE-ViT-Base model is pre-trained on FSD50K and stopped either at the same wall-clock time as our method or after 50 epochs.
The third involves reported representative models from prior work, including supervised large-scale pre-training and SSL models pre-trained on AudioSet or LibriSpeech.
All experiments are implemented in PyTorch and run on a single NVIDIA RTX 4090. Pre-training uses a ViT-Base backbone with $16{\times}16$ patches on $1024{\times}128$ log-Mel inputs, masking ratio 75\%, batch size 64, and AdamW\cite{loshchilov2017decoupled} for 100 epochs (for complete architectural details and pre-training hyperparameter configurations, please refer to Table S1 in the supplementary material).
Downstream fine-tuning uses unmasked inputs with batch size 32, and SpecAugment\cite{park2019specaugment} is applied.
For the primary benchmark evaluations reported in Table~\ref{tab:overall_results}, models are fine-tuned for 300 epochs to ensure complete convergence.


\subsection{Core Results on ESC-50 and UrbanSound8K}
\label{ssec:core_results}

Table~\ref{tab:overall_results} summarizes fine-tuning performance across ESC-50, US8K, SCv2, and FSD50K.
Under full fine-tuning, AudioPG reaches 90.60\% on ESC-50 and 88.17\% on US8K, compared to 54.00\% and 75.34\% for training from scratch.
These gains are also reflected under frozen-feature evaluation. On ESC-50 linear probing, AudioPG improves over a raw log-Mel baseline.
Since both baselines use the same log-Mel representation, this comparison isolates the benefit of the learned encoder under a simple classifier.
We also analyze misclassification patterns across multiple evaluation datasets to characterize the limitations of representations learned from procedural pre-training.
Table~\ref{tab:error_analysis} summarizes the most frequent confusions and their underlying acoustic similarities (an extended cross-dataset error analysis is provided in Table S2 of the supplementary material).
The errors predominantly reflect physical proximity in the learned representation space where fine-tuning fails to fully separate classes with overlapping time and frequency structures.
\begin{table*}[t]
 \captionsetup{skip=2.5pt}
\centering
\caption{Cross-Dataset Error Analysis and Acoustic Attribute Attribution.}
\label{tab:error_analysis}
\fontsize{9pt}{10.8pt}\selectfont
\setlength{\aboverulesep}{0pt}
\setlength{\belowrulesep}{0pt}
\renewcommand{\arraystretch}{1.1}
\begin{tabularx}{\textwidth}{lllc *{4}{>{\centering\arraybackslash}X}}
\toprule
\multirow{2}{*}{\textbf{Dataset}} & \multirow{2}{*}{\textbf{True Class}} & \multirow{2}{*}{\textbf{Confused As}} & \textbf{Rate} & \multicolumn{4}{c}{\textbf{Primary Error Source (Acoustic / Semantic)}} \\ \cmidrule(l){5-8}
 & & & \textbf{(\%)} & \textbf{Transient} & \textbf{Mechanical} & \textbf{Broadband} & \textbf{Semantic/Phonetic} \\ \midrule
\multirow{5}{*}{ESC-50} 
 & Footsteps & Fireworks & 75.0 & \ding{51} & & & \\
 & Helicopter & Engine & 37.5 & & \ding{51} & & \\
 & Helicopter & Rain & 37.5 & & & \ding{51} & \\
 & Vacuum cleaner & Washing machine & 37.5 & & \ding{51} & \ding{51} & \\
 & Breathing & Door wood creaks & 25.0 & 
\ding{51} & & \ding{51} & \\ \midrule
\multirow{3}{*}{US8K} 
 & Air conditioner & Siren & 11.0 & & \ding{51} & & \\
 & Siren & Dog bark & 10.8 & \ding{51} & \ding{51} & & \\
 & Air conditioner & Street music & 10.0 & & & \ding{51} & \ding{51} \\ \midrule
\multirow{3}{*}{Speech} 
 & Forward & Four & 15.8 & & & & \ding{51} \\
 & Tree & Three & 13.8 & \ding{51} & & & \ding{51} \\
 & Off & Up & 7.2 & \ding{51} & & \ding{51} & \\ \midrule
\multirow{5}{*}{FSD50K} 
 & Mechanical fan & Vehicle & 90.0 & 
& \ding{51} & & \\
 & Mechanical fan & Engine & 80.0 & & \ding{51} & & \\
 & Tearing & Domestic sounds & 78.6 & \ding{51} & & & \ding{51} \\
 & Microwave oven & Engine & 76.2 & & \ding{51} & & \\
 & Camera & Domestic sounds & 73.9 & \ding{51} & & & \ding{51} \\ \bottomrule
\end{tabularx}

\vspace{0.25em}
\begin{minipage}{\textwidth}
\footnotesize \textit{Note:} 
\textbf{Transient}: Impulsive broadband bursts or rhythmic onsets.
\textbf{Mechanical}: Continuous low-frequency rumble or periodic amplitude modulation. 
\textbf{Broadband}: Non-stationary stochastic noise textures.
\textbf{Semantic / Phonetic}: Broad ontology overlap or human vocal tract phonetic similarities.
\end{minipage}
\end{table*}

Several recurring confusions involve transient overlap and broadband noise textures.
In the ESC-50 dataset, footsteps are frequently confused with fireworks due to shared impulsive broadband onsets and rhythmic temporal spacing.
Similarly, the FSD50K dataset exhibits severe confusions between shatter and crushing events.
The transient injection module in our procedural curriculum produces comparable vertical structures in the spectrogram, which makes it difficult for the model to separate short impulsive events without additional dataset-specific contextual supervision.
Another prominent error category involves sustained sources with amplitude modulation or mechanical resonance.
Helicopter sounds are confused with engines, and vacuum cleaners are confused with washing machines.
In FSD50K, mechanical fans are frequently misclassified as vehicles or engines.
These errors align with the physical synthesis process where oscillatory or pulse excitations combined with low-pass damping produce similar continuous low-frequency rumble and periodic modulation patterns across different semantic categories.
In the Speech Commands dataset, errors are strongly driven by phonetic overlap rather than environmental noise.
Words including forward and four, or tree and three, share dominant vowel formants and initial consonant transient properties.
Since the procedural generator does not explicitly model human vocal tract formants or complex linguistic articulation, the pre-trained representation relies on generic spectral shapes which are insufficient to resolve fine-grained phonetic distinctions.
Overall, a substantial fraction of these errors corresponds to acoustically similar pairs, indicating that fine-tuning primarily reshapes boundaries but struggles to eliminate physical ambiguities inherent to the representation.
\subsection{Ablation Study and Efficiency Analysis}
\label{ssec:ablation_efficiency}

To quantify how individual synthesizer components contribute to downstream transfer, and to compare procedural pre-training against real-data pre-training under matched compute budgets, we conduct a comprehensive analysis on ESC-50 and US8K.
Note that to isolate the effects of pre-training efficiency and scale, all downstream fine-tuning in this subsection strictly employs a Fast Evaluation Protocol constrained to a 50-epoch fine-tuning schedule.
This isolates component contributions efficiently but results in lower absolute performance compared to the 300-epoch schedule used for our main benchmark evaluations.
Results are summarized in Table~\ref{tab:ablation_efficiency}.

\begin{table*}[t]
 \captionsetup{skip=2.5pt}
\centering
\caption{Comprehensive Ablation and Efficiency Analysis.}
\label{tab:ablation_efficiency}
\fontsize{9pt}{10.8pt}\selectfont
\setlength{\aboverulesep}{0pt}
\setlength{\belowrulesep}{0pt}
\renewcommand{\arraystretch}{1.05}
\setlength{\tabcolsep}{7.5pt}
\begin{tabularx}{0.95\textwidth}{Xcccccccc}
\toprule
\multirow{2}{*}{\textbf{Configuration}} & \textbf{Freq.} & \textbf{Timbre} & \textbf{Dyn.} & \textbf{Filter} & \textbf{Real Data} & \textbf{PT Time} & \multicolumn{2}{c}{\textbf{Accuracy (\%)}} \\ \cmidrule(l){8-9}
 & ($f_0$) & (Harm.) & (ADSR) & (Damp.) & \textbf{(hours)} & \textbf{(sec)} & \textbf{ESC-50↑} & \textbf{US8K↑} \\ \midrule
\multicolumn{9}{l}{\quad\textit{I.
Synthetic Ablation (50 epochs)}} \\
1. Sine (Baseline) & \checkmark & -- & -- & -- & -- & 668 & 42.25 & 70.97 \\
2. Harmonic & \checkmark & \checkmark & -- & -- & -- & 819 & 69.25 & 75.99 \\
3. Dynamic & \checkmark & \checkmark & \checkmark & -- & -- & 847 & 73.25 & 78.73 \\
4. AudioPG (Ours) & \checkmark & \checkmark & \checkmark & \checkmark & -- & 1123 & \textbf{82.00} & \textbf{85.75} \\ \midrule
\multicolumn{9}{l}{\quad\textit{II.
High Resource Regime (Long-term Convergence)}} \\
AudioPG (Ours, 500 ep) & \checkmark & \checkmark & \checkmark & \checkmark & -- & $\sim$11,000 & 85.40 & 88.17 \\
Real (Full Training) & -- & -- & -- & -- & $\sim$100 h & 13,044 & \textbf{87.00} & \textbf{88.38} \\ \midrule
\multicolumn{9}{l}{\quad\textit{III.
Real Data Reference (Time-Matched / Cold Start)}} \\
Real (Time-Matched) & -- & -- & -- & -- & $\sim$100 h & 1,100 & 72.00 & 83.57 \\ \bottomrule
\end{tabularx}%

\vspace{0.25em}
\begin{minipage}{0.95\textwidth}
\footnotesize \textit{Note:} We evaluate the contribution of each synthesizer component and compare procedural pre-training against real-data pre-training under aligned computational budgets.
Filter refers to spectral damping simulation.
\end{minipage}
\end{table*}

\textbf{Component ablations.}
Rows 1 to 4 of Table~\ref{tab:ablation_efficiency} progressively increase synthesizer complexity.
The frequency-only baseline reaches 42.25\% on ESC-50 and 70.97\% on US8K. Adding harmonic structure improves performance to 69.25\% and 75.99\%;
adding temporal dynamics further improves to 73.25\% and 78.73\%. The full AudioPG configuration, which additionally includes spectral damping and background noise, achieves 82.00\% on ESC-50 and 85.75\% on US8K.
The relatively strong US8K performance of the frequency-only baseline is consistent with stationary mechanical classes dominated by narrowband structure, whereas ESC-50 benefits more from added timbral and temporal diversity.
\begin{figure}[h!]
  \centering
  \includegraphics[width=\columnwidth]{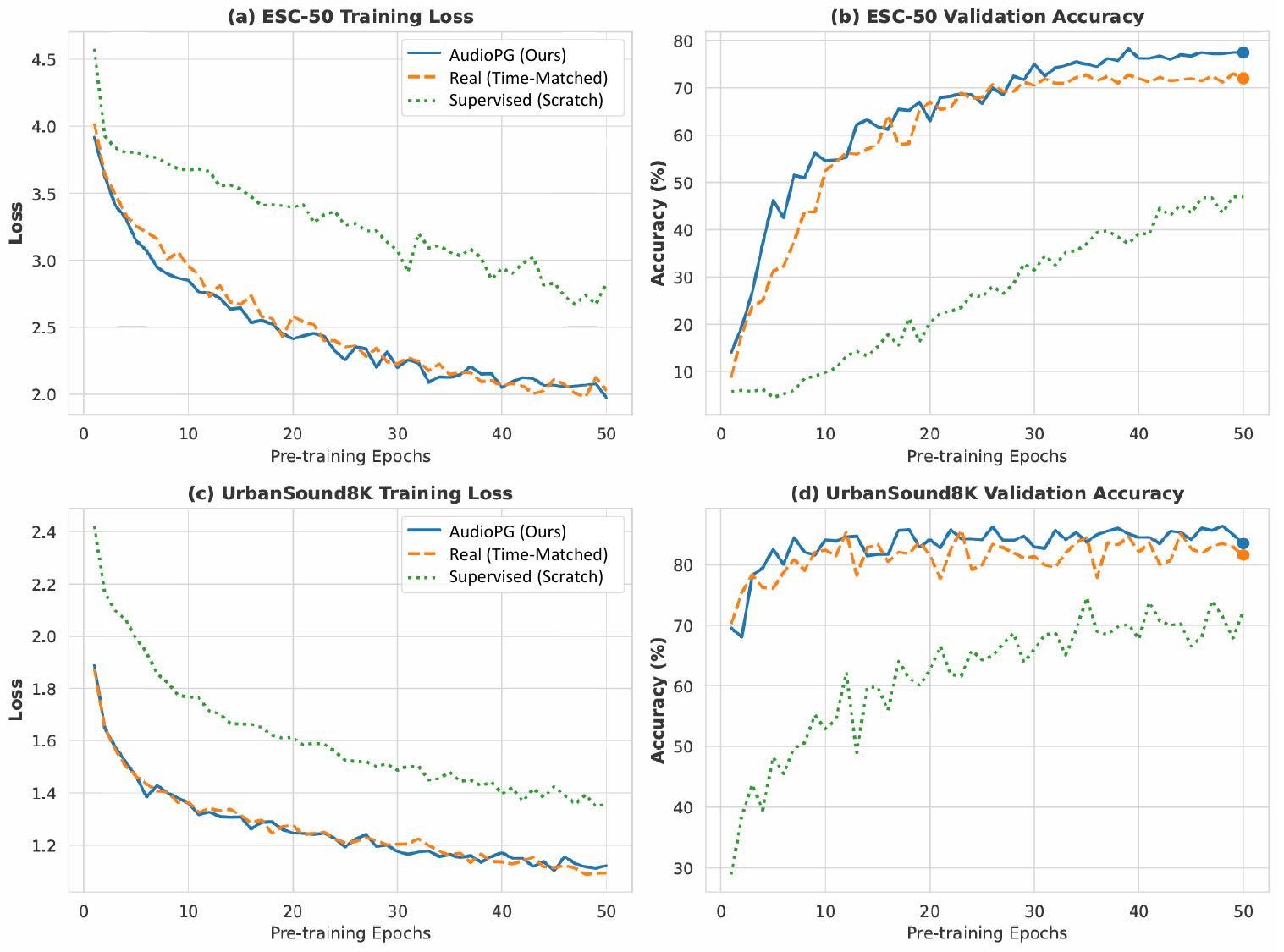} 
  \caption{\textbf{Cold-Start Efficiency Dynamics.} 
  Validation accuracy and training loss on ESC-50 and UrbanSound8K over the first 50 epochs.
AudioPG converges faster than the time-matched real-data baseline under the same wall-clock budget.}
  \label{fig:efficiency}
\end{figure}

\textbf{Cold-start efficiency.}
Under a strict time budget, AudioPG achieves 82.00\% on ESC-50, compared to 72.00\% for the time-matched real-data baseline.
Fig.~\ref{fig:efficiency} shows that the gap emerges early and persists throughout the budget.
One possible contributing factor is that on-the-fly synthesis reduces the prevalence of near-silent segments and provides consistently structured time and frequency patterns for reconstruction, which can be beneficial under constrained compute.
\textbf{Scaling analysis and long-term convergence.}
To examine scalability, we evaluate checkpoints at multiple pre-training stages.
Accuracy increases rapidly in the early phase and continues to improve with longer training.
In Table~\ref{tab:ablation_efficiency}, extending procedural pre-training to 500 epochs improves ESC-50 accuracy to 85.40\%.
In the high-resource regime, the real-data baseline reaches 87.00\%, suggesting a remaining domain gap between procedural and real audio that becomes more visible as compute increases.
\begin{figure}[t]

  \centering
  \includegraphics[width=\columnwidth]{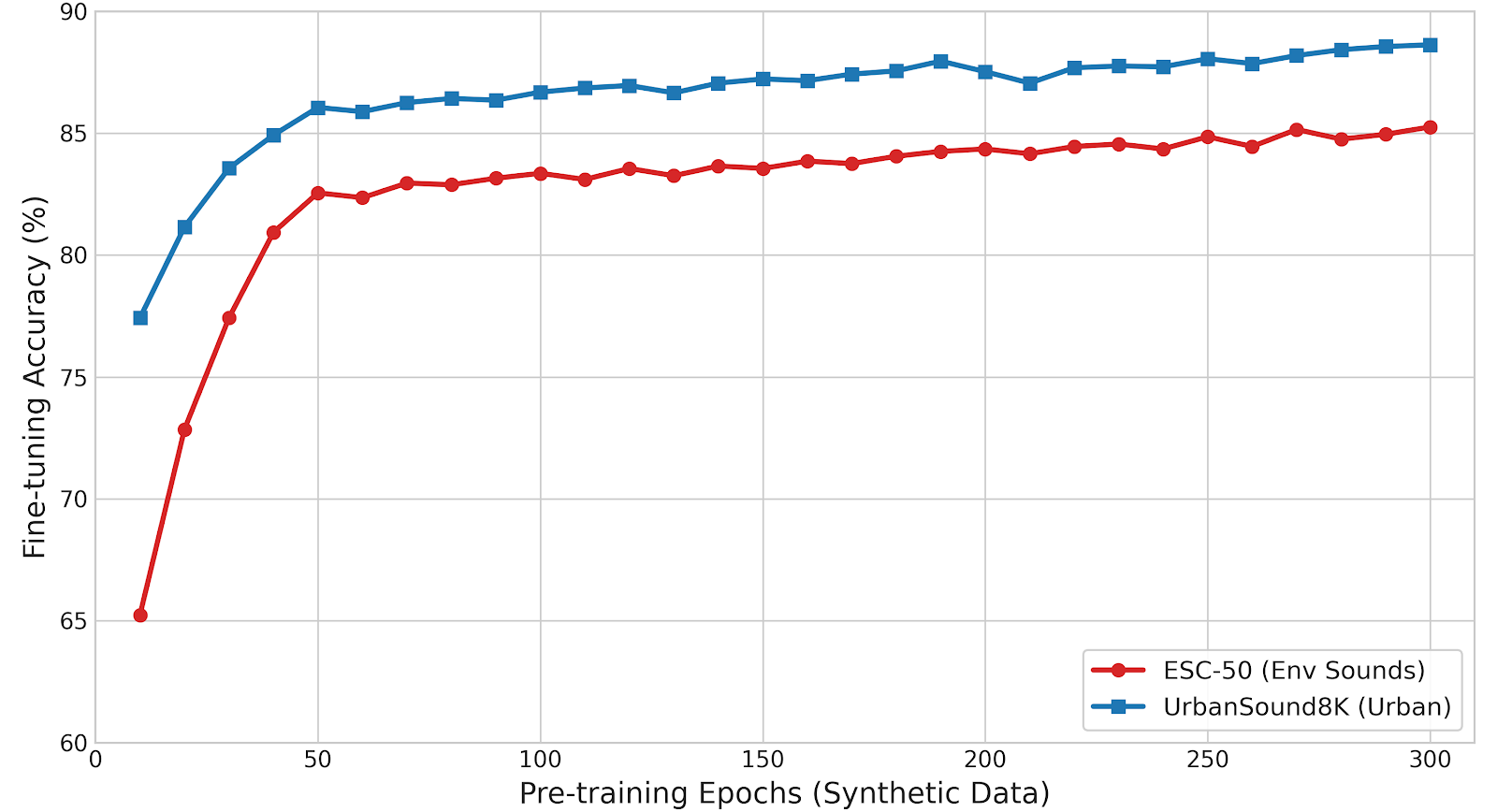} 
  \caption{\textbf{Scaling Behavior.} Downstream accuracy on ESC-50 as a function of pre-training epochs.
Evaluated under the fast 50-epoch fine-tuning protocol, the model exhibits rapid learning in the early phase and steady refinement in the long term.}
  \label{fig:scaling}
\end{figure}

\begin{table}[h!]
 \captionsetup{skip=2.5pt}
\centering
\caption{Sensitivity Analysis of Physical Parameters.}
\label{tab:sensitivity}
\fontsize{9pt}{10.8pt}\selectfont
\setlength{\aboverulesep}{0pt}
\setlength{\belowrulesep}{0pt}
\renewcommand{\arraystretch}{1.05}
\setlength{\tabcolsep}{4.5pt}
\begin{tabularx}{0.95\columnwidth}{cXcc}
\toprule
\textbf{No.} & \textbf{Physical Configuration} & \textbf{US8K↑ (\%)} & \textbf{ESC-50↑ (\%)} \\ \midrule
\multicolumn{4}{l}{\quad\textit{Frequency Bandwidth ($f_0$ Range)}} \\
1 & Narrow ($50\sim500$ Hz) & 84.95 & 76.50 \\
2 & Ours ($50\sim2000$ Hz) & \textbf{86.14} & 76.75 \\
3 & Wide ($50\sim4000$ Hz) & 86.02 & \textbf{77.00} \\ \midrule
\multicolumn{4}{l}{\quad\textit{Event Density (Events per clip)}} \\
4 & Sparse (1 Event) & 81.24 & 70.75 \\
5 & Ours (1--5 Events) & \textbf{86.14} & 76.75 \\
6 & Crowded (5--10 Events) & 
83.51 & \textbf{78.75} \\ \bottomrule
\end{tabularx}%
\vspace{0.25em}
\begin{minipage}{0.95\columnwidth}
\footnotesize \textit{Note:} The ours configuration denotes the default setting used in AudioPG.
\end{minipage}
\end{table}

\textbf{Sensitivity to physical parameters.}
We vary the fundamental frequency range and event density (Table~\ref{tab:sensitivity}).
Widening the frequency range yields small changes on ESC-50, while event density has a larger effect, where increasing from 1 event to multiple events improves ESC-50 accuracy.
This indicates that overlapping events can provide a more challenging reconstruction signal that benefits downstream transfer.
\begin{figure*}[htbp]
\captionsetup{skip=0.5em}
  \centering
  \includegraphics[width=0.85\textwidth]{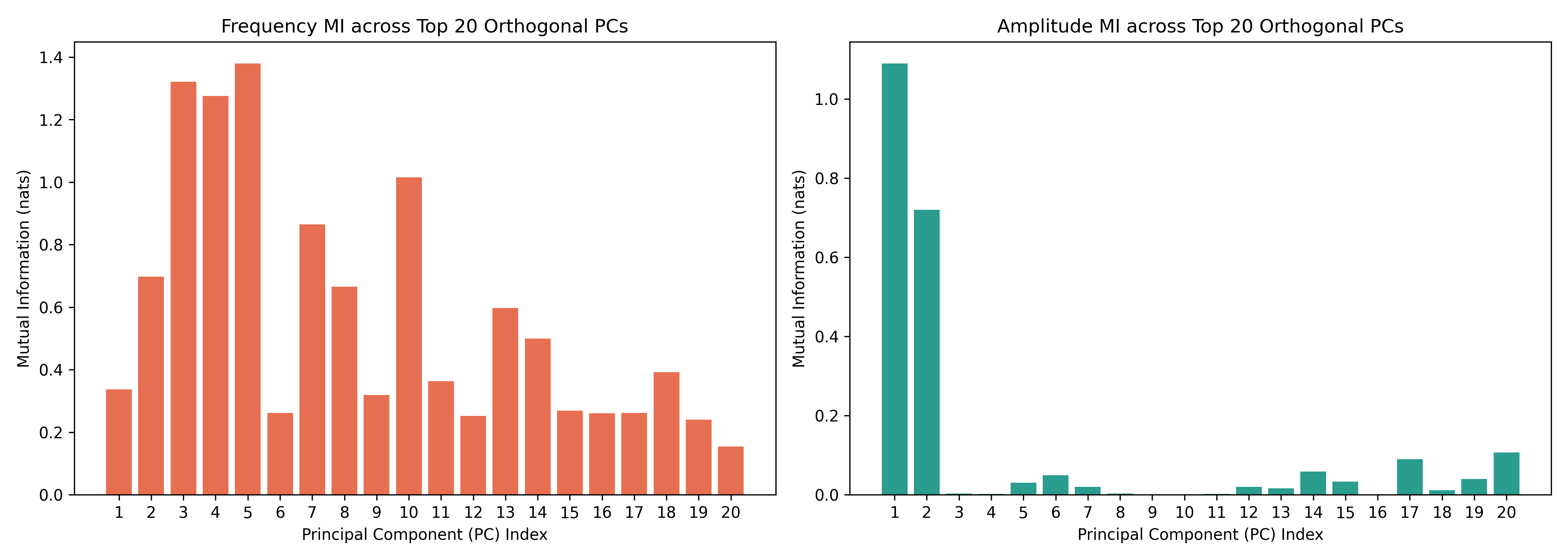}
  \caption{
  \textbf{Orthogonal Subspace Disentanglement.} Mutual Information spectrum across the top 20 Principal Components of the frozen embeddings.
The right panel shows that amplitude dominates the primary variance directions.
The left panel shows that frequency information is separated into orthogonal subspaces, indicating that the model disentangles physical properties at the manifold level.}
  \label{fig:pca_spectrum}
\end{figure*}
\subsection{Emergent Disentanglement of Physical Factors}
\label{ssec:neuron_analysis}

We hypothesize that reconstructing procedurally generated audio induces latent representations tracking generative factors. We analyze frozen encoder embeddings against ground truth physical parameters including frequency, relative intensity, and temporal position using mutual information, linear probing, and principal component analysis. Using the Kraskov k-Nearest Neighbor estimator\cite{kraskov2004estimating}, we quantify mutual information between individual latent dimensions and continuous physical factors (details in supplementary material Section S-I.C and Fig. S2). Specific neurons strongly correlate with physical factors, where Neuron \#343 tracks frequency with a mutual information of 1.7299, Neuron \#666 varies monotonically with relative intensity at 1.4660, and Neuron \#437 correlates with temporal position at 0.4896. Low Mutual Information Gaps across frequency (0.0062), amplitude (0.1273), and temporal position (0.1277) indicate the model adopts a distributed representation where redundant dimensions jointly encode physical properties. A Ridge regression probe on frozen embeddings achieves high R-squared scores for frequency (0.9946), amplitude (0.9916), and temporal position (0.9550), demonstrating projection into a linearly decodable physical manifold. Principal component analysis reveals the top 20 orthogonal components explain 96.49\% of total latent variance. As illustrated in Fig.~\ref{fig:pca_spectrum}, physical factors disentangle at the subspace level. Amplitude is primarily captured by the first principal component with a mutual information of 1.0343, while frequency information separates into subsequent orthogonal subspaces peaking at the fifth component with a mutual information of 1.4124. This confirms the latent space variance directions align with independent physical generative factors.
\subsection{Evolution of Acoustic Filters During Pre-training}
\label{ssec:filter_evolution}

\begin{figure*}[t]
  \centering
  \includegraphics[width=\textwidth]{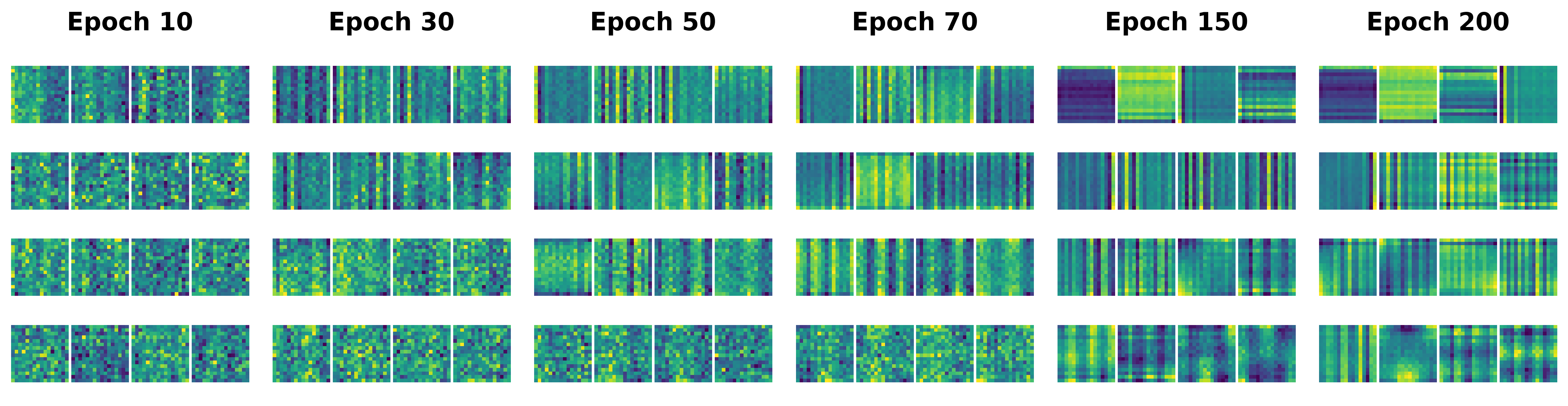}
  \caption{\textbf{Evolution of Patch Embedding Filters.}
  Visualization of the top PCA components of the patch embedding filters throughout training.
Early epochs show noisy and unstructured components; mid training exhibits emerging time and frequency patterns;
later epochs show clearer horizontal selectivity bands and vertical transient-like edges.}
  \label{fig:filter_evolution}
\end{figure*}

To visualize how the model internalizes time and frequency structure during pre-training, we analyze the evolution of the first-layer patch projection weights, which map spectrogram patches into the initial token embeddings.
We apply principal component analysis to flattened patch filters at selected training epochs and visualize the leading components inFig.~\ref{fig:filter_evolution}.
Across training, the components evolve from noise-like patterns to more structured time and frequency templates.
In later epochs, several components resemble frequency-selective bands that align with harmonic stacks and onset-sensitive edges that align with transient bursts and envelope boundaries present in the procedural curriculum.
This analysis complements the neuron-level results by providing a view of how low-level patch projections evolve as the masked reconstruction objective is optimized.
\subsection{Preliminary Exploration on Target-Domain Adaptation}
\label{sec:preliminary_dapt}

We briefly explore target domain adaptation on the unlabeled FSD50K dataset. We emphasize that this preliminary exploration does not constitute a primary contribution of our current work, but serves to demonstrate potential applications that future work will follow up on. As shown in Table~\ref{tab:dapt}, using AudioPG initialization for continued masked autoencoding reaches 91.80\% accuracy on UrbanSound8K, 90.80\% on ESC-50, and 0.617 mAP on FSD50K. This outperforms the ImageNet baseline and models pretrained on massive real world corpora including AudioSet on aligned environmental domains. Furthermore, unlike the ImageNet initialized model which suffers from domain bias and degrades to 88.25\% on ESC-50, AudioPG preserves performance on broader downstream tasks. This suggests procedural priors provide a robust starting point against performance degradation when adapting to limited real world datasets. Detailed investigations into these domain adaptive paradigms are deferred to future research.

\begin{table}[t]
\captionsetup{skip=2.5pt}
\caption{Evaluation of Target-Domain Adaptation}
\label{tab:dapt}
\centering
\resizebox{\linewidth}{!}{
\begin{tabular}{llccc}
\toprule
\textbf{Method / Initialization} & \textbf{Pre-training Data} & \textbf{US8K} ($\uparrow$) & \textbf{ESC-50} ($\uparrow$) & \textbf{FSD50K} (mAP $\uparrow$) \\
\midrule
MAE-AST [2] & AS + LS ($\sim$2M) & 88.40\% & 90.00\% & 0.482 \\
MaskSpec [51] & AudioSet ($\sim$2M) & 89.60\% & 90.70\% & 0.503 \\
MSM-MAE [37] & AudioSet ($\sim$2M) & 89.80\% & \textbf{94.00\%} & 0.490 \\
\midrule
ImageNet Init & FSD50K ($\sim$50K) & 89.96\% & 88.25\% & 0.569 \\
\textbf{AudioPG Init (Ours)} & \textbf{FSD50K} ($\sim$50K) & \textbf{91.80\%} & 90.80\% & \textbf{0.617} \\
\bottomrule
\end{tabular}
}
\end{table}

\vspace{-6pt}
\section{Discussion}
\label{sec:discussion}

AudioPG demonstrates that a parameterized procedural generator effectively replaces fixed real-audio corpora for representation learning. The controlled synthesis signal yields transferable representations and enables latent space analyses difficult with uncontrolled real-world data.
\vspace{-6pt}
\subsection{Emergence of Physically Grounded Features}
\label{ssec:disc_mechanism}

Model transferability stems from the masked reconstruction objective capturing shared time and frequency regularities. Patch-embedding visualizations reveal an evolution from unstructured noise to frequency-selective bands and transient-like edges aligning with the synthesis process and real environmental audio, qualitatively corroborated by reconstruction analysis on held-out samples in the supplementary material. Furthermore, while individual neurons exhibit distributed encoding, principal component analysis and linear probing confirm the model projects complex acoustic signals into a structured manifold. Physical attributes including global intensity and spectral properties separate into orthogonal and linearly decodable subspaces, indicating procedural pre-training induces features anchored to basic acoustic structures without explicit supervision.
\vspace{-9pt}
\subsection{Algorithmic Parsimony and Semantic Limitations}
\label{ssec:disc_limitations}

Strong multimedia audio performance typically requires large-scale curated corpora, imposing access and storage constraints. AudioPG alleviates these bottlenecks via a lightweight, zero-real-audio pipeline. Unlike systems requiring iterative clustering or contrastive optimization, AudioPG utilizes a standard masked autoencoder and a parameterized generator producing physically plausible structures. Acoustic primitives provide a more relevant pre-training signal than non-acoustic frameworks like visual fractals.

Despite this efficiency, physics-driven pre-training presents a semantic gap, as physical similarity inconsistently aligns with human semantics. Error analysis shows acoustically plausible confusions persist after fine-tuning because the generator lacks high-level semantics, complicating semantic-heavy classifications. This necessitates richer source models and compositional priors to reflect real-world semantics while preserving controllable generation.

To address this representational divergence, future research can integrate large-scale generative models like AudioGen~\cite{kreuk2022audiogen} and latent diffusion architectures~\cite{liu2023audioldm,copet2024simple} as high-level rule generators. Instead of expensive acoustic rendering, foundation models can output structural configurations like temporal distributions and modulation parameters. Mapping abstract concepts to synthesizer states enables semantic-aware procedural curricula at negligible cost. Ultimately, fusing cognitive priors with acoustic primitives bridges physical principles and human understanding, enhancing generalization across multimedia applications.
\vspace{-6pt}
\section{Conclusion}
\label{sec:conclusion}
We presented AudioPG, a procedural synthesis framework for audio representation learning that requires no real recordings. AudioPG combines a dynamic synthesizer with a Transformer masked autoencoder to reconstruct log Mel spectrogram patches. The encoder transfers to real audio benchmarks through fine-tuning, achieving 90.60\% accuracy on ESC-50, 88.17\% on UrbanSound8K, 0.546 mAP on FSD50K, and 97.03\% on Speech Commands V2. Pre-training completes in less than 20 minutes on a single GPU. Latent analysis shows the model disentangles physical factors including frequency, intensity, and onset into orthogonal and linearly decodable subspaces. Patch embedding projections also develop structured time-frequency patterns during the training process. These results suggest that controllable procedural generators provide an effective pre-training signal when large real audio corpora are unavailable. Future work will expand the generator to match real-world diversity and mixing conditions while maintaining control and synthesis efficiency.

\begin{acks}
This work was supported in part by Participation in Research Program of Shanghai Jiao Tong University(Grant No. T541PRP49003) ,the Startup Fund for Young Faculty at SJTU (SFYF at SJTU, Grant No. 25X010506040) and the National Undergraduate Training Program for Innovation and Entrepreneurship under Grant 202610269116G.
\end{acks}

\newpage
\bibliographystyle{ACM-Reference-Format}
\bibliography{refs}

\end{document}